\documentstyle[aps,prd,floats,graphicx,twocolumn]{revtex}
\def\beq{\begin{equation}}
\def\eeq{\end{equation}}
\def\bea{\begin{eqnarray}}
\def\eea{\end{eqnarray}}

\def\al{\alpha}

\newcommand{\square}{\kern1pt\vbox{\hrule height  1.2pt\hbox{\vrule
width 1.2pt\hskip 3pt \vbox{\vskip 6pt}\hskip  3pt\vrule width
0.6pt}\hrule height 0.6pt}\kern1pt}

\begin{document}

\twocolumn[\hsize\textwidth\columnwidth\hsize\csname
@twocolumnfalse\endcsname

\title{Mapping the Dark Energy with Varying Alpha}

\author{David Parkinson$^{\dag}$, Bruce A. Bassett$^{\dag}$ and John D. Barrow$^{*}$}

\address{$^{\dag}$Institute of Cosmology and Gravitation,  University of Portsmouth, Portsmouth~PO1~2EG, UK}
\address{$^{*}$ DAMTP, Centre for Mathematical Sciences, University of Cambridge, Cambridge, CB3 0WA, UK}

\date{\today}
\maketitle

\begin{abstract}
Cosmological dark energy is a natural source of variation of the fine structure constant. Using a model-independent
approach we show that once general assumptions about the alpha-varying interactions are made, astronomical probes of its variation constrain the  dark energy equation of state today to satisfy $-1 < w_f < -0.96$ at 3-sigma and significantly disfavour late-time changes in the equation of state.
We show how dark-energy-induced spatial perturbations of alpha are linked to violations of the Equivalence Principle and are thus negligible at low-redshift, in stark contrast to the 
BSBM theories. This provides a new test of dark energy as the source of alpha variation.
\end{abstract}
\vskip 3em
]

The universe appears to be accelerating in response to a dark energy whose
identity remains a mystery. We show how the current astronomical data
supporting a $5\sigma $ detection of time variation in the fine-structure
`constant', $\alpha =e^{2}/\hbar c,$ at redshifts $0.2<z<3.7$ \cite%
{qso1,qso2} can provide direct evidence distinguishing a dynamical dark
energy from a cosmological constant, which is very challenging for standard
cosmological tests, e.g. \cite{cmb}. Unlike cosmological tests which depend on
the integrated properties of the dark energy, measurement of $%
\frac{\Delta \alpha(z)}{\alpha} \equiv \frac{\alpha(z)-\alpha(0)}{\alpha(0)}$ would
allow us to map the dark energy as a function of $z$, see also refs.\cite%
{chiba,history,gold}.

\underline{{\em Mapping the dark energy }} -- We will consider a variation
of $\alpha $ generated by a non-trivial gauge kinetic function (GKF), 
$Z_{F}(\phi )$, leading to an electromagnetic lagrangian 
$\frac{1}{4}Z_{F}(\phi )F_{\mu \nu }F^{\mu \nu}$, where $\phi $ is
some scalar field and $F_{\mu \nu }$ the usual Maxwell tensor. (We
expect $Z_{F}\neq 1$ in string theory through the coupling to the
dilaton \cite{fede} or through non-renormalisable interactions in
supergravity \cite{carroll}) In this case $\alpha (z)\propto
Z_{F}(\phi )^{-1}$ so if we know $Z_{F}$ and $\Delta\alpha(z) /\alpha$
precisely we can deduce $\phi (z)$. 

Is it natural for $\phi $ to be the dark energy? The interaction $Z_{F}(\phi
)F^{2}$ is not renormalisable in 4d and hence $Z_{F}=Z_{F}(\phi /M)$
where $M$ is the mass scale at which the effective theory description (usually
supergravity) breaks down. If we assume $M$ is the Planck mass, $M_{pl}\sim
10^{19}GeV$, only very large expectation values of $\phi $ will lead to
measurable effects. This is what happens in many dark-energy models that
typically require large field values, $\phi /M_{pl}=O(1)$ today to satisfy
the slow-roll condition, $M_{pl}^{2}V_{\phi \phi }<V,$ on the
potential $V(\phi )$ needed for an accelerating universe. We will
show, using a model-independent approach, how one can constrain the
dark-energy's equation of state. Note that such models for varying
$\alpha $ differ from those of Bekenstein-Sandvik-Barrow-Magueijo
(BSBM) \cite{BSBM} in which $\alpha $-variations are driven by
coupling to the charged non-relativistic matter
alone and so the associated scalar field cannot be the dark energy.

We can also consider spatial variation of $\alpha $ \cite{BM,qso2}. Dark
energy should cluster on scales larger than its Compton wavelength $\lambda
_{c}\propto V_{\phi \phi }^{-1/2}$, which exceeds $100Mpc$ in standard
quintessence models \cite{quint}, although in more exotic models  this
need not be true\cite{condens}. A map of the {\em spatial variation}
of $\alpha $ can potentially probe both the power spectrum and Compton
wavelength of fluctuations in the dark energy. Consider the spatial
fluctuation $\delta \alpha \equiv \alpha (x_{\mu })-\overline{\alpha }$.
Quantities with an overbar are evaluated in the background cosmology
at the same event. To first order in $\delta \phi $ we have 
\begin{equation}
\frac{\delta \alpha ({\bf x},t)}{\alpha }=(\partial _{\phi }\ln \alpha |_{%
\overline{\phi }})~\delta \phi \,.  \label{fluc}
\end{equation}%
Evaluated today, this is directly related to Weak Equivalence Principle
(WEP) constraints \cite{wep} on varying-$\alpha $ theories which, assuming
large $\lambda _{c}$ so that we may use the background solution for $\phi $
even in the nonlinear regime\footnote{By contrast, in BSBM
  varying-$\alpha $ theories spatial variation in $\alpha $ is strong
  because it is driven by gradients in electromagnetically-charged
  matter, removing the power of local constraints \cite{MB}.}, 
imply $\delta \alpha /\alpha |_{0}\leq 10^{-5}\delta \phi /M_{pl},$
today. Since we expect $\delta \phi \ll M_{pl}$, spatial fluctuations
of $\alpha $ should be negligible at low $z$ if dark energy (or any
light scalar field) is the source for the time-variation of $\alpha
$. This is in agreement with current data \cite{qso2}. But at high $z$
the spatial fluctuations of $\alpha $ could be
significant. Eq. (\ref{fluc}) implies that the power spectrum of
$\alpha $ fluctuations, $P_{\delta\alpha }(k)\equiv \frac{k^{3}}{2\pi
  ^{2}}|\delta \alpha _{k}|^{2}$ is anti-correlated with the spectrum
of dark-energy fluctuations, $P_{\delta \phi}(k)$. Their ratio, which
we call the {\em $\alpha $-bias} $b_{\alpha }$, will in general be
time-dependent with potentially interesting implications for the CMB
\cite{cmb2}.

\begin{figure}[tbp]
\includegraphics[width=90mm]{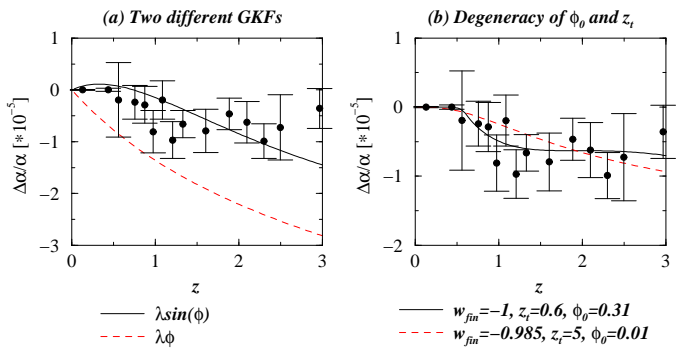} \\
\caption[1dplots]{Degeneracies: (a) $\Delta \protect\alpha(z)
  /\protect\alpha$ for the same
dark-energy model with two different choices of $Z_{F}$. Note that $Z_{F}=%
\protect\lambda \sin \protect\phi $ allows both positive and negative values
of $\Delta \protect\alpha(z) /\protect\alpha$. (b) The $z_{t}-\protect\phi %
_{0}/M$ degeneracy in fitting the 
$\Delta \protect\alpha(z) /\protect\alpha$  data. Both models have almost the same $\protect\chi ^{2}$ value. The WEP
 data breaks this degeneracy. }
\label{user}
\end{figure}

\underline{{\em $Z_{F}$ degeneracy }} -- Our discussion has been predicated
on the (unreasonable) assumption that we know the function $Z_{F}$. A
standard approach \cite{history} is to assume slow roll for $\phi $ in some
potential $V(\phi )$ and expand $Z_{F}$ around $\phi _{0}$, the value of $%
\phi $ today (see for example ref. \cite{gold}). The leading term gives $%
\Delta \alpha(z) /\alpha \propto \Delta {\phi }/M_{pl}$. However this is
overly restrictive since $Z_{F}$ may not be analytic in $\phi $, the dark
energy may not derive from a slowly-rolling scalar field \cite{PR}, or $Z_{F}
$ may change very rapidly with $\phi $, e.g. $Z_{F}\propto e^{\phi }$ so
even if $\phi \sim \ln (t)$, $\Delta \alpha(z) /\alpha$ would exhibit
power-law variation with redshift. Hence there is a perfect degeneracy
between $Z_{F}$ and the dark-energy dynamics, shown in Fig (\ref{user}a) by
plotting $\Delta \alpha(z) /\alpha$ for the {\em same} dark-energy model
with two very different GKFs: $Z_{F}(\phi )=\lambda \phi $ and $\lambda \sin
(\phi )$. If the resulting curves were assumed to arise from the same GKF,
they would have been interpreted as coming from two completely different
dark-energy models.

To further illustrate this degeneracy, consider the left branch of the Oklo
analysis of the neutron capture resonance \cite{oklo} which suggests $\Delta
\alpha /\alpha =(0.88\pm 0.07)\times 10^{-7}$. This corresponds to a larger $%
\alpha $ at $z=0.13$, in contrast to the Webb {\em et al} data which
strongly suggests $\alpha $ was smaller at $z>0.5$ with $\Delta \alpha
/\alpha =-(0.543\pm 0.116)\times 10^{-5},$ \cite{qso2}.

It has been claimed that this is inconsistent with a slowly-rolling
quintessence origin; true if $Z_{F}$ is monotonic in $\phi $. However, a GKF
such as $Z_{F}(\phi )=\lambda \sin (\phi )$ can give both positive and
negative $\frac{\Delta \alpha(z) }{\alpha }$ and match the apparently
inconsistent data with monotonic $\phi (z)$, as shown in Fig. 1(a). In fact,
given any data $\Delta \alpha(z) /\alpha$ and any smooth $\phi (z)$ we can
reconstruct a corresponding GKF which will {\em exactly} fit the data by
defining the GKF to be $Z_{F}(\phi (z))=\alpha _{\ast }/\alpha (z)-1$ where $%
\alpha _{\ast }\simeq (1+Z_{F}(\phi _{0}))/137$ normalises $\alpha (z)$.

Dark energy can induce variation of $\alpha $ in other ways. We can choose a
tensor, $g_{EM}^{\mu \nu }$, to raise and lower indices on $F_{\mu \nu }$
that differs from the spacetime metric $g^{\mu \nu }$ used to form the Ricci
scalar $R$ \cite{BLMV}. A simple example is 
\begin{equation}
g_{EM}^{\mu \nu }=g^{\mu \nu }\left( 1+\frac{(\nabla \phi )^{n}}{M^{2n+2}}%
\nabla ^{\mu }\phi \nabla ^{\nu }\phi \right) ,
\end{equation}%
where $\nabla ^{\mu }$ is the covariant derivative w.r.t. $g^{\mu \nu }$.
This yields $Z_{F}=Z_{F}(|\nabla _{\mu }\phi |/M^{2})$, so $\Delta \alpha
/\alpha \sim (H/M)$. If $M=M_{pl}$ such interactions are negligible today;
but if $M\sim 10^{-3}eV$, the models can easily fit the data at 
$z\leq 3.5$; however, if $\phi $ is a tracker field then $\dot{\phi}%
\rightarrow \infty $ as $z\rightarrow \infty $. If $Z_{F}(\phi )\propto 1+(%
\dot{\phi}/M^{2})^{n}$ for any $n$, then $|Z_{F}(\infty )|\rightarrow \infty 
$ and $\alpha (z)\rightarrow 0$ rapidly. In fact, we have found that models
of this form that match the data at $z<3$ but strongly violate the
nucleosynthesis constraint $|\frac{\Delta \alpha }{\alpha }(10^{10})|<0.1$ 
\cite{robl}. This can be avoided by choosing a GKF with $Z_{F}(\dot{\phi}\rightarrow \infty )=$ constant, or 
a model with bounded $|\dot{\phi}|$, as in condensation models of dark
energy \cite{condens}, but this goes beyond the scope of the paper.\footnote{%
Another example, the 1-loop QED lagrangian in curved spacetime, \cite%
{drumhath} yields a GKF of the form 
\begin{equation}
Z_{F}=1+aR^{2}+bR_{\mu \nu }R^{\mu \nu }+cR_{\mu \nu \gamma \delta }R^{\mu
\nu \gamma \delta }\,,
\end{equation}%
where $a,b,c$ are constants and $R^{\mu \nu \gamma \delta }$ is the Riemann
tensor. However, these terms are Planck suppressed and are not expected to
be influential today.}. This shows the importance of theoretical input into
choosing $Z_{F}$ and the need for general parametrisations of $Z_{F}$. Below
we investigate to what extent we can expect to distinguish between different 
$Z_{F}$ and different dark energy equations of state.

\underline{{\em Consistency}} -- Before we discuss mapping of the dark
energy we address in more detail the suggestions in the literature (e.g. 
\cite{chiba,string}) that it is difficult, or even impossible, for a
slowly-rolling scalar field to match both null geonuclear data \cite{oklo,olive1} 
at $z<0.5$ and the quasar data at $z>0.5$ in $Z_{F}$ theories (it is
easier in BSBM theories because there is strong spatial variation \cite{MB}%
). To quantify the extent to which dark energy can match the data we use
model-independent approach \cite{condens,meta,CC,linder} which uses $%
w(z)\equiv p_{\phi }/\rho _{\phi }$ instead of choosing a particular
potential. An expedient choice is%
\begin{equation}
w(z)=w_{f}[1+\exp (((z-z_{t})/\Delta )]^{-1}\,,  \label{w}
\end{equation}%
so that $w(z\rightarrow \infty )=0$; $w_{f}$ is constant; $z_{t}$ is the
redshift at which a transition to dark-energy domination occurs, while the
constant $\Delta $ controls its width. This form accurately describes most
quintessence models and some exotic alternatives such as vacuum
metamorphosis \cite{PR}. We fix $\Delta =z_{t}/30$ as in \cite{meta,condens}
so $w(0)=w_{f}$, and perform a likelihood analysis over the free dark-energy
parameters $(w_{f},z_{t})$ and the parameters of the GKF.
\begin{figure}[tbp]\begin{center}
\includegraphics[width=60mm]{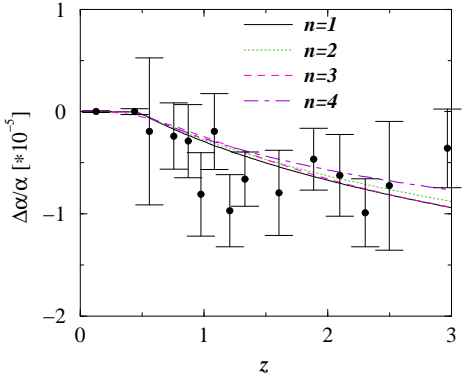} \\
\caption[1dplots]{The individual best-fits for each value of $n$. The
difference in $\Delta \protect\alpha /\protect\alpha $  between the models for $z\leq3$ is
less than $2\times 10^{-6}$ and yet Figure (\protect\ref{vacombined}) shows
that the combined data strongly favour $n=3$. This reflects the highly
non-Gaussian nature of the $\protect\chi ^{2}$ hypersurfaces. }
\label{bestfits}
\end{center}\end{figure}
We have already seen that if $Z_{F}$ is completely free one can match {\em %
any} data with {\em any} dark-energy dynamics. Therefore, we need
theoretical input or a general parametrisation of the GKF. In the spirit of
the model-independent approach we consider an expansion of the GKF in the
form: 
\begin{equation}
Z_{F}=1-\Gamma _{n}\varphi ^{n}\,,~~\Gamma _{n}\equiv \beta _{n}\left( \frac{%
\phi _{0}}{M}\right) ^{n}  \label{zf}
\end{equation}%
where $\varphi (z)\equiv \phi (z)/\phi _{0}$ and $\Gamma _{n}>0$ are
dimensionless couplings. We limit ourselves to $n\leq 4$ for computational
reasons and consider each $n$ separately. This choice of $Z_{F}(\phi )$
expands our set of parameters to $(w_{f},z_{t};\beta ,\frac{\phi _{0}}{M},n)$. 
The best-fits for each $n$, varying the remaining parameters, are shown in 
Fig. (\ref{bestfits}).

\begin{figure}[tbp]
\includegraphics[width=80mm, height=80mm]{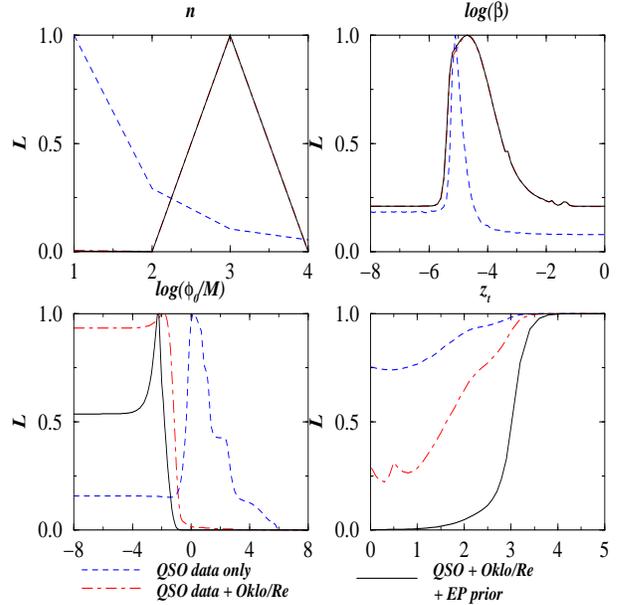} \\
\caption[1dplots]{The marginalised 1d-likelihoods for $n,\protect\beta ,%
\protect\phi _{0}/M$ and $z_{t}$ for each of the three combinations of data
sets. The curves represent the final likelihoods including all the data.
Note how the data strongly favour $n=3$ and how the WEP data selects $%
\protect\phi _{0}/M\simeq 10^{-2}$ as the most favoured value. }
\label{vacombined}
\end{figure}


Our first $\Delta \alpha(z) /\alpha$ data set is \{QSO\}, the quasar
absorption data \cite{qso1,qso2} providing the evidence for time-variation
of $\alpha $. The second set of data is \{QSO + Oklo + Re\}. \{Oklo\} is the
null branch of the Oklo data ($|\frac{\Delta \alpha }{\alpha }(z=0.13)|\leq
10^{-7}$) \cite{oklo}; \{Re\} is the from the Os/Re ratio from radioactive
decay $^{187}%
\mathop{\rm Re}%
\rightarrow ^{187}Os$ in iron meteorites, \cite{olive1} ($|\frac{\Delta
\alpha }{\alpha }(z=0.44)|<3\times 10^{-7}$) close to the lowest redshift
quasars. Last, we add composition-dependent WEP constraints on the
difference in the acceleration, $a$, of different bodies $A$ and $B$, $\eta
\equiv (\Delta a/a)_{AB}\leq 10^{-12}$ \cite{wep,chiba} that arises
because of their different numbers of charged nucleons. Allowing for gauge
coupling unification, $\eta \propto (d\ln \alpha /d\phi )^{2}$ \cite{fede}
(and note eq. (\ref{fluc})) and this translates into the bound
\begin{equation}
\frac{n\Gamma _{n}}{(1-\Gamma _{n})}\left[ \frac{M}{\phi _{0}}\right]
<10^{-5}\frac{M}{M_{pl}}\,,  \label{wep}
\end{equation}%
today. 

The best-fit reduced-$\chi ^{2}$ for the three data sets are $4.7/8$, $%
4.8/10$ and $10.4/11$ respectively. Adding the WEP constraint clearly has
the largest effect but the $\chi ^{2}$ values show that general dark-energy
models are compatible with the current $\al(z)$ data. 

\underline{{\em Closing the circle}} -- Is it possible to gain information
about $Z_{F}(\phi )$ and the dark energy simultaneously? To answer this
question we compute the 1-d likelihoods for each one of the parameters in
the set $(w_{f},z_{t};\beta ,\frac{\phi _{0}}{M},n)$ by marginalising over
the remaining parameters on a grid consisting of $2.3$ million points using
the following precise expression for $\Delta \alpha(z) /\alpha$: 
\begin{equation}
\frac{\Delta \alpha (z)}{\alpha }=-\Gamma _{n}\frac{1-\varphi ^{n}(z)}{%
1-\Gamma _{n}\varphi ^{n}(z)}\,.  \label{delalpha}
\end{equation}%
Our results, are shown in Fig. (\ref{vacombined}).

In particular, a dimensionless coupling $\beta \sim 10^{-3}-10^{-5}$ is
preferred by all the data. The role of $n$ is interesting. While for each $n$
it is possible to find sets of parameters with nearly equal $\chi ^{2}$, Fig. (\ref{bestfits}), 
the marginalised likelihood for $n$ is selective, Fig.(\ref{vacombined}). 
This reflects the fact that the likelihood hypersurface is highly non-Gaussian, 
with a large number of nearly degenerate minima. However the
most likely values of our parameters depend on the data set used, casting
some doubt on the maturity of current data. Considering only the \{QSO\}data
set, $n=1$ is preferred. However, when the low-redshift \{Oklo/Re\} and WEP
constraints are added then $n=3$ is significantly favoured, which is a
promising sign for future determination of the GKF. If we consider only the
\{QSO + Oklo + Re\}data, there is an interesting degeneracy between
dark-energy models with low (high) $z_{t}$ and high (low) values of $\phi
_{0}/M$. This is illustrated in Fig. (\ref{user}b) by two good fits to the
data with parameter values $(z_{t},\phi _{0}/M)=(5.0,10^{-2})$ and $%
(0.6,10^{0.5})$ respectively. 

\begin{figure}[h]
\begin{center}
\includegraphics[width=50mm]{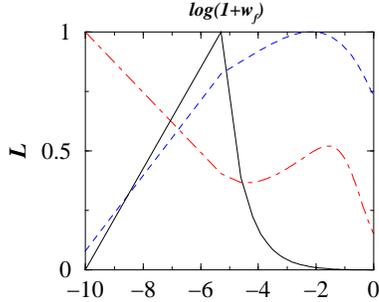} \\
\caption[1dplot]{The 1-d likelihood for $\log (1+w_{f})$. The dashed line
corresponds to the quasar data, the dot-dashed line to the quasar+Oklo+Re-Os
data, and the solid line to all the data (including the WEP violation
constraint). The full data set gives $-1<w_{f}<-0.96$ at $3\protect\sigma $. 
}
\label{wf}
\end{center}
\end{figure}

In order to map the dark energy we calculate
 $\phi(z) = \phi_0 - \int_0^z \dot{\phi}(z) dz/(H(1+z))$ where 
$\dot{\phi}^{2}(z) =(1+w(z))\rho _{\phi }$ and $\rho _{\phi }(z)=\rho _{0}\exp (3\int
(1+w)dz/(1+z))$\cite{meta,condens}.  We set $\Omega _{\phi }=0.7$ today. These equations imply
that as $w_{f}\rightarrow -1$ and $\Delta \alpha /\alpha \rightarrow 0$ for $%
z<z_{t}$. Clearly, the tighter the null results on $\Delta \alpha $ at low
redshift become, the closer $w_{f}$ is pushed to $-1$. This can be clearly
seen in the likelihood curve for $w_{f}$ in Fig. (\ref{wf}) when the \{Oklo +Re\} and WEP data
are added. The WEP constraint is particularly useful since it breaks the
degeneracy between \ $\beta ,\phi _{0}$ and $M$ in eq. (\ref{delalpha})
which only depends on the combination $\Gamma _{n}$. The constraint is
weakest if we choose $M=M_{pl}$, as we do.

The complete set of data favour both $w_{f}$ close to $-1$ and $%
z_{t}>2.5$; see Figs (\ref{wf}) \& (\ref{vacombined}). \{Oklo + Re\} favours 
$\dot{\phi}(z<0.45)=0$ and hence $w_{f}\simeq -1$. Adding WEP restricts $-1<w_{f}<-0.96$ at $%
3\sigma $ with $w_{f}=-1$ significantly disfavoured, see Fig. (\ref{wf}). This is because $%
w_{f}=-1$ would imply $\Delta \alpha(z) /\alpha =0$ for $z<z_{t},$ and to match
the quasar data would then require $z_{t}<0.5$, inconsistent with the
likelihood curve for $z_{t}$. Small values of $z_{t}$ values are disfavoured
because, as shown in Fig. (\ref{user}), there is a $z_{t}-(\phi _{0}/M)$
degeneracy in matching the quasar data. Models with small $z_{t}$ require
large values of $\phi _{0}/M$, but large $\phi _{0}/M$ is disfavoured by the
Oklo/Re data (Fig \ref{vacombined}). Conversely, very small $\phi _{0}/M$
are disfavoured (at constant $\Gamma _{n}$) by the WEP data, eq. (\ref{wep}).

For $n=1$ (the standard dimension-5 case \cite{chiba,history}) the
constraint (\ref{wep}) yields $\beta _{1}<10^{-5}(M/M_{pl}),$ while for $%
n\geq 2$ it is a joint constraint on $\Gamma _{n}$ and $\phi _{0}/M$. In the
next decade the STEP \cite{step}, GG \cite{gg} and MICROSCOPE \cite{mic}
experiments promise sensitivities up to $\eta <10^{-18}$ and hence a null
detection would imply $d\ln \alpha /d\phi <10^{-8}$ today, virtually ruling
out $n=1$. To investigate potential future constraints we can impose this
expected bound on our best-fits for each $n$. The resulting reduced-$\chi
^{2}$s all exceed $38/11$ and would all be excluded by this
future data.

In summary, by using a model-independent approach to the dark energy, we
have shown that the current constraints on $\Delta \alpha(z) /\alpha$ can
be well-matched by dark-energy models. For a large family of varying-$\alpha 
$ theories, we have shown how spatial fluctuations in $\alpha $ are
correlated with fluctuations of the dark energy but need to be negligibly
small at low-redshifts due to Weak Equivalence Principle constraints, so
providing a clear test of the scenario. We have pointed out that the gauge
kinetic function, $Z_{F}$, responsible for the variation of $\alpha $ is
completely degenerate with the dark energy dynamics so that any dynamical
dark-energy model can be made to fit the $\frac{\Delta \alpha(z) }{\alpha}$
data. Nevertheless, with a reasonable parametrisation of $Z_{F}$ the current
data already yield interesting constraints on $Z_{F}$. When the WEP
constraint is added the current equation of state is forced to lie in the
range $-1<w_{f}<-0.96$ at $3\sigma $, the tightest constraints on the dark
energy yet found. This is unlikely to change much if one varies other cosmic parameters.
Late-time change in $w(z)$ at $z < 2$ and $\frac{\phi_0}{M}>10^{-1}$ are also strongly disfavoured, 
in contrast to standard CMB/LSS results \cite{meta,condens}, raising the possibility that a 
complete combined analysis will rule out most dark-energy models or even perhaps the paradigm of 
dark-energy-induced variation of $\al$. 

Finally, we note
that the GKFs we have considered break the conformal invariance of the
Maxwell equations and may thus allow the seeding of magnetic fields during
inflation. This would provide an interesting way of linking the early universe to
observations in the present epoch.

We thank John Webb for supplying us with the QSO data, and Federico
Piazza, Rob Crittenden, Haim Goldberg, Sujata Gupta, Martin Kunz, Andrew Liddle,  
Andr\'{e} Lukas and Michael Murphy for useful discussions and comments.

\end{document}